\documentclass{elsart}
\input{epsf}

\begin{document}
\begin{frontmatter}
\title{Tricritical behavior of random systems with a coupling to a
nonfluctuating parameter}

\author{Y.N.Skryabin\thanksref{NSM}}
\address{Institute for Metal Physics, Russian Academy of Sciences,\\
Ural Division, Kovalevskaya Str., 18,\\
620219, Ekaterinburg, Russia \\}
\author{A.V.Shchanov}
\address{Ural State Technical University, \\
Mir Str., 19,
620002, Ekaterinburg, Russia}
\thanks[NSM]{Partially supported by State Program ``Actual Problems in 
Condensed Matter Physics: Neutron Studies''(Projects No. 96-104, 96-305) 
and Russian Foundation for Basic Research (Project No. 97-02-17315), 
Russia.}

\begin{abstract}
The influence of disordering upon critical behavior of the system with 
hidden degrees of freedom is considered. It is shown that there is a 
tricritical behavior in the constrained system, while in the unconstrained 
system only phase transitions of the second order occur.
\end{abstract}
\end{frontmatter}

\section{Introduction}
Recently, phase transitions under a constraint imposed on some ``hidden''
degrees of freedom coupled to the critical ones have again attracted
attention \cite{lyra}. As it is well known, many real systems possess a hidden
degree of freedom \cite{nelson,kosterlitz,chaves}, which can be described
as the nonfluctuating parameter in the vicinity of the phase transition.
Fisher \cite{fisher} stressed the importance of the constraint upon the
hidden degree of freedom. Fisher's theory of phase transition in constrained
systems was generalized \cite{imry,dohm} to include the possibility of first-
order phase transitions. In particular, it was found that the tricritical
behavior coincides with the critical behavior of the ideal system provided
$\alpha>0$ ($\alpha$ is the critical exponent of the specific heat) and with
the renormalized one provided $\alpha<0$. Early renormalization group studies 
of systems with constrained nonfluctuating parameter were presented in Refs.
\cite{chaves,achiam1,achiam2}.

As it is shown in Ref.\cite{lyra}, the nature of a phase transition may
strongly depend on the constraint imposed on the system. For the case of the
unconstrained system the stability criterion for the occurrence of continuous
phase transition can be found within the mean field theory. On the other
hand, for the case of a constrained system a different criterion, which is
distinct from the mean-field one, is calculated. This criterion results in a
fluctuation-induced renormalized Heisenberg tricritical point. 

The main purpose of this paper is to consider the influence of disordering
upon critical behavior of a system with hidden degrees of freedom and to
focus attention on a criterion for the occurrence of continuous phase
transition in this system.
Such disordering can be caused, e.g., by the presence of
``frosen'' impurities \cite{lubensky}. The analysis of renormalization group
equations for random system with constrained nonfluctuating parameter has been
presented in Ref.\cite{ls}. It was found the fixed points and shortly
described the critical behavior. 
Here we discuss the tricritical behavior and especially a first-order 
transition taking into account results of \cite{lyra}.

\section{Renormalization group equations}
Let us consider the disordered system in which an $n$-component vector order
parameter ${\bf S}({\bf x})$ is coupled with the scalar nonfluctuating order
parameter $y({\bf x})$. This system can be described in momentum space by the 
effective Hamiltonian \cite{ls}
\begin{eqnarray}
H &=& \frac12 \int \frac{d{\bf q}_1}{(2\pi)^d}\frac{d{\bf q}_2}
{(2\pi)^d}r({\bf q}_1, {\bf q}_2) {\bf S}({\bf q}_1) {\bf S}({\bf q}_2)
+\nonumber\\
& &+\int \frac{d{\bf q}_1}{(2\pi)^d}\frac{d{\bf q}_2}{(2\pi)^d}
\frac{d{\bf q}_3}{(2\pi)^d}\frac{d{\bf q}_4}{(2\pi)^d}u({\bf q}_1, {\bf q}_2,
{\bf q}_3, {\bf q}_4){\bf S}({\bf q}_1) {\bf S}({\bf q}_2){\bf S}({\bf q}_3)
{\bf S}({\bf q}_4)+\nonumber\\
& &+\int \frac{d{\bf q}_1}{(2\pi)^d}\frac{d{\bf q}_2}{(2\pi)^d}
\frac{d{\bf q}_3}{(2\pi)^d} \mu({\bf q}_1, {\bf q}_2, {\bf q}_3)y({\bf q}_1)
{\bf S}({\bf q}_2) {\bf S}({\bf q}_3)+\nonumber\\
& &+\frac12 \int \frac{d{\bf q}_1}{(2\pi)^d}\frac{d{\bf q}_2}{(2\pi)^d}
\beta({\bf q}_1, {\bf q}_2)y({\bf q}_1)y({\bf q}_2)+
\int \frac{d{\bf q}}{(2\pi)^d}h({\bf q})y({\bf q}).
\label{hamiltonian}
\end{eqnarray}

The Hamiltonian (\ref{hamiltonian}) can be used to study the ferromagnetic
phase transition in the three-dimensional Hubbard model with frosen
impurities. Herewith, the coupling $\mu$ is pure imaginary
\begin{equation}
\mu=i(kT_c)^{1/2}U^{3/2}N'(E_f),
\label{coupling}
\end{equation}
where $N(E_f)$ is the density of states on the Fermi energy $E_f$ and $U$ is
the Coulomb potential. The study of this case is of particular interest.

Following the standard method of the renormalization group \cite{wilson} and 
its
extension to random systems \cite{lubensky} the recurrence relations for the
potential averages over the probability distribution function can be
constructed by averaging on the recurrence relations for the potentials of
the Hamiltonian (\ref{hamiltonian}) of the inhomogenious system and for the
second cumulants:
\begin{eqnarray}
r' &=& b^2\{r+[4(n+2)\tilde u-\tilde\Delta+2nz]A(r)\}, \label{r}\\
\tilde u' &=& b^{\epsilon}\{\tilde u-K_4 \ln b [4(n+8){\tilde u}^2-
6\tilde u\tilde\Delta]\}, \label{u}\\
\tilde\Delta' &=& b^\epsilon\{\tilde\Delta-K_4 \ln b [8(n+2)\tilde u\tilde
\Delta-4{\tilde\Delta}^2]\}, \label{delta}\\
z' &=& b^\epsilon\{z-K_4 \ln b [8(n+2)\tilde uz+2nz^2-2z\tilde\Delta]\},
\label{z}\\
w'&=& b^\epsilon\{w-K_4 \ln b [8(n+2)\tilde uw+2nw^2+4zw-2w\tilde\Delta]\},
\label{w}
\end{eqnarray}
where we introduce the following symbols for second cumulants:
$\Delta\sim<rr>_c$,
$\nu\sim<rh>_c$, $\rho\sim<hh>_c$, and use the notation:  $\tilde u=u-
\frac{\mu^2}{2\beta}$, $\tilde\Delta=\Delta-\frac{4\mu\nu}{\beta}+
\frac{4\mu^2\rho}{\beta^2}$, $z=\frac{\mu^2}{\beta}-\frac{\mu_0^2}{\beta_0}$,
$w=\frac{\mu_0^2}{\beta_0}$.
All other definitions are standard in renormalization group theory:
$K_4=\frac{1}{8\pi^2}$ is a quality proportional to the surface area of the
unit sphere in $d=4$ space,
$b$ is the scale parameter and $A(r)$ is the integral of a closed loop
of pair correlation functions (see, also, \cite{is}).

In Eqs.(\ref{r}-\ref{w}) we separate the coefficient of the nonfluctuating
parameter $y({\bf q}=0)$ from those of $y({\bf q}\not=0)$ because of their
possible role in constraining systems \cite{chaves,achiam1}. We make a shift
of the variable $y({\bf q}=0)$ for the disappearance of the linear term in
the variable $y$ on each step of the renormalization procedure. It is also
necessary to write the equations for the cumulants
$\nu$, $\nu_0$, $\rho\mu$, $\rho_0\mu_0$. However, the structure of the
new equations is such that they do not violate the stability of fixed points,
which were defined by Eqs.(\ref{r}-\ref{w}). Using the definition of
cumulants $\Delta$, $\nu$, $\rho$  one can show that $\tilde\Delta\geq0$.

\begin{figure}
\epsfbox{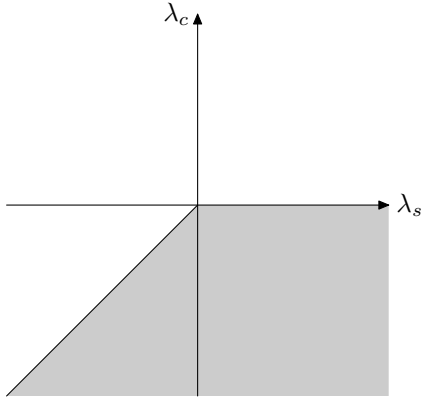}
\caption{Phase diagram in the mean-field approximation for the
unconstrained system}
\label{fig:1}
\end{figure}

Before discussing the renormalization group analysis, consider the mean-field
theory results. After integrating over the nonfluctuating parameter in
(\ref{hamiltonian}), we obtain a new effective Hamiltonian for order parameter
${\bf S}({\bf q})$. A solution of this new integrated out effective
Hamiltonian in the mean-field approximation gives the boundary conditions
of instability for the Hubbard model:
$\lambda_s-\lambda_c^{(0)}=0$, $\lambda_c^{(0)}=0$, where $\lambda_s\equiv u$
and $\lambda_c^{(0)}=\mu_0^2/2\beta_0$,
for unconstrained system ($\lambda_c^{(0)}=\lambda_c^{(1)}=\lambda$)
and $\lambda_s=0$, $\lambda_c^{(1)}=0$, where $\lambda_c^{(1)}=\mu^2/2\beta$,
for the
constrained system ($\lambda_c^{(0)}=0$). These conditions divide the first
and second order phase transition ranges in
($\lambda_c,\lambda_c^{(1)}$)-$\lambda_s$--plane
(Fig. \ref{fig:1} and Fig. \ref{fig:2}, see, also, \cite{lyra}).
\begin{figure}
\epsfbox{fig.2}
\caption{Phase diagram in the mean-field approximation for the constrained
system}
\label{fig:2}
\end{figure}
Using the Stoner criterion $r_s\equiv1-(U/2)N(E_f)=0$, we have the condition
for a continuous phase transition for the unconstrained system
\begin{equation}
\frac{3(N'(E_f))^2}{2N(E_f)}>N''(E_f),
\label{unconstraint}
\end{equation}
while the first order phase transition occurs for the contrary case. For
the constrained system we have the continuous phase transition, if
\begin{equation}
N''(E_f)<0.
\label{constraint}
\end{equation}
According to the mean-field theory for random systems, the
impurities do not influence on the stability ranges of the system.

A feature of the recurrence relations is a closed system of two equations:
(\ref{u}) and (\ref{delta}). It is easy to find the fixed points of this
system. We have as usual the Gaussian fixed point ($G$) with
$\tilde u^{\ast}=\tilde \Delta^{\ast}=0$, the Heisenberg fixed point ($H$)
with $\tilde u^{\ast}=\epsilon/4K_4(n+8)$, $\tilde\Delta^{\ast}=0$,
the non-physical fixed point with $\tilde u^{\ast}=0$,
$\tilde \Delta^{\ast}=-\epsilon/4K_4$, and random fixed point ($R$)
with $\tilde u^{\ast}=\epsilon/16K_4(n-1)$,
$\tilde\Delta^{\ast}=\epsilon(4-n)/8K_4(n-1)$.
Using these fixed points, one can easy find the rest fixed points from the
recurrence relations.

\begin{figure}
\epsfbox{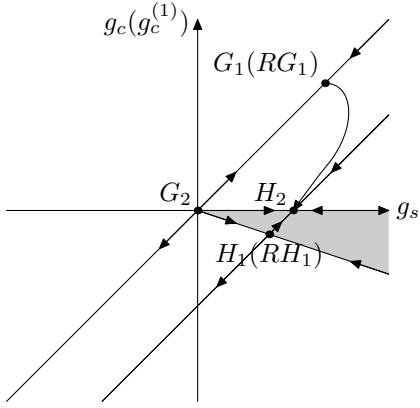}
\caption{Flow trajectories in the $\tilde\Delta=0$ plane}
\label{fig:3}
\end{figure}

\begin{figure}
\epsfbox{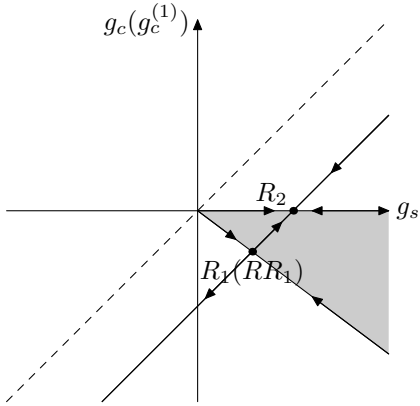}
\caption{Flow trajectories in the $\tilde\Delta=\tilde\Delta^{\ast}$ plane}
\label{fig:4}
\end{figure}

The full list of fixed points, the critical behavior, which is determined
by their values, and their stability have been presented in \cite{ls}.
We can see from the Table in \cite{ls} that some parameters and their
selfvalues $\lambda_i$ change 
sign at $n=4$. This changing in sign strongly influences the forms of
flow trajectories, and we are obliged further to discuss
separately the $n<4$ and $n>4$ cases.
It is convenient to use the two parameter spaces: ($\lambda_s,\lambda_c,
\tilde\Delta$) for the
unconstrained case ($\lambda_c^{(0)}=\lambda_c^{(1)}=\lambda_c$) and
($\lambda_s,\lambda_c^{(1)},\tilde\Delta$) for the constrained case
($\lambda_c^{(0)}=0$). Moreover, it is turn out that the interesting 
fixed points are in the planes 
$\tilde\Delta=0$ and $\tilde\Delta=\tilde\Delta^{\ast}\not=0$ in both
spaces. Due to the relation $\tilde u=u-\mu^2/2\beta\equiv \lambda_s-
\lambda_c^{(1)}$ all fixed points align
on two parallel lines in these planes.
The flow trajectories are identical in both spaces, so we can
demonstrate them simultaneously in the $\tilde\Delta=0$ (Fig. \ref{fig:3})
and $\tilde\Delta=\tilde\Delta^{\ast}$ (Fig. \ref{fig:4}) planes in the spaces
of the renormalized coupling constants $g_s$, $g_c^{(0)}$, and $g_c^{(1)}$.
The notation of fixed points, which correspond to the subspace
($\lambda_s,\lambda_c^{(1)},\tilde\Delta$) are written in brackets.
It is important to point out that fixed points with
$z^{\ast}\equiv2\left(\lambda_c^{(1)^\ast}-\lambda_c^{(0)^\ast}\right)\not=0$
determine the renormalized critical behavior in accordance with \cite{fisher}.

One can see from Fig. \ref{fig:3} and Fig. \ref{fig:4} that the most stable
fixed points and, hence, the critical behavior is independent of the existence
of constraint. One point is $H_2$, when $n>4$, and the critical behavior is
ideal for a pure system and another point is $R_2$, when $n<4$, and the
critical behavior is determined by the presence of the frosen inhomogeneities.
Thus, the behavior coincides with the one found in \cite{lubensky}.

The critical behavior at the boundary of the stability region is determined by
the fixed points $R_1$ (for $n<4$) and $H_1$ (for $n>4$) in the 
system without
constraints, and by the ``renormalized'' fixed points $RR_1$ (for $n<4$) and
$RH_1$ (for $n>4$) in system with constraints.
It is easy to see that the boundary of the stability region in constrained
systems reduces the criterion (\ref{constraint}) for pure system to
\begin{equation}
\frac{3(N'(E_f))^2}{2N(E_f)}<\frac{\alpha_H/n\nu_H}
{\epsilon/(n+8)+\alpha_H/n\nu_H}N''(E_f)+O(\epsilon),
\label{pureconstraint}
\end{equation}
where $\alpha_H=\frac{4-n}{2(n+8)}\epsilon$ and
$\nu_H=\frac12+\frac{n+2}{4(n+8)}\epsilon$
are the specific heat and the correlation length critical exponents
for the Heisenberg fixed point, respectively (see, also, \cite{lyra}).
By analogy with a pure system, one can find the corresponding criterion for 
a random constrained system in the form
\begin{equation}
\frac{3(N'(E_f))^2}{2N(E_f)}<\frac{\alpha_R/n\nu_R}
{\epsilon/4(n-1)+\alpha_R/n\nu_R}N''(E_f) +O(\epsilon),
\label{randomconstraint}
\end{equation}
where $\alpha_R=\frac{n-4}{8(n-1)}\epsilon$ and
$\nu_R=\frac12+\frac{3n}{32(n-1)}\epsilon$ are corresponding exponents for
the random fixed point.
Thus, in the constrained system the flow trajectories run away in both the pure
system fixed point $H_1$ and the random fixed point $R_1$, and, hence, the
critical behavior at the boundary of the stability region reduces to the
tricritical behavior.

\section{The tricritical behavior}
Using the standard field theory and renormalization group method (see,
for example,
\cite{amit}), we obtained the next equation for the free energy in the
one loop approximation
\begin{eqnarray}
\beta F(T,M) &=& \frac12r_s(T)M^2+\frac{1}{4!}(\lambda_s-
                                    \lambda_c^{(0)})M^4-\nonumber\\
& &-\frac12\tilde\Delta M^2\int\frac{d{\bf q}_1}{r_s(T)+q^2}-
          \frac{6}{4!}\lambda_2\tilde\Delta M^4\int\frac{d{\bf q}_1}
                                        {(r_s(T)+q^2)^2}+\nonumber\\
& &+\frac12(n-1)\int d{\bf q}_1\ln\Bigl(1+\frac{\lambda_1M^2/6}
                                      {r_s(T)+q^2}\Bigr)+\nonumber\\
& &+\frac12\int d{\bf q}_1\ln\Bigl(1+\frac{\lambda_2M^2/6}{r_s(T)+q^2}\Bigr),
\label{frenerg}
\end{eqnarray}
where $M=\Bigl(\sum_{\alpha}M_\alpha^2\Bigr)^{1/2}$ is the magnetization, and
\begin{equation}
\lambda_1=\frac13(\lambda_s-\lambda_c^{(0)}),\quad
\lambda_2=\frac13(3\lambda_s-\lambda_c^{(0)}-2\lambda_c^{(1)}).
\label{couplings}
\end{equation}
Note that here we used coefficients of Eq.(\ref{hamiltonian}) in
form: $u=\lambda_s /4!$, $\mu_0^2/2\beta_0=\lambda_c^{(0)}/4!$,
$\mu^2/2\beta=\lambda_c^{(1)}/4!$.
To get the renormalized free energy, we used the
theory of Refs.\cite{amit} and \cite{lyra}. Finally, we have
\begin{eqnarray}
\beta F(t,M)-\beta F(t,0) &=& \frac12\tilde tM^2+\frac{1}{4!}(\tilde g_s-
\tilde g_c^{(0)})M^4+\nonumber\\
& &(n-1)f(t+\frac12g_1 M^2)+f(t+\frac12g_2 M^2),
\label{frenergy}
\end{eqnarray}
where $\tilde t$, $\tilde g_s$, and $\tilde g_c^{(0)}$ are the
renormalized by frozen impurities values, $g_1$ and $g_2$ are renormalized
coupling constants $\lambda_1$ and $\lambda_2$ and
\begin{equation}
f(x)=\frac18x^2(\ln x - \frac12).
\end{equation}
For the unconstrained case $g_c^{(0)}=g_c^{(1)}=g_c$ we have the conventional
$\phi^4$--model with an effective coupling constant proportional to $g_s-g_c$
and the runaway flow trajectories do not intersect the boundary of the 
stability range. Hence, we can suggest that the phase transition remains 
second order.
It should be noted here, that the fixed points $H_1$ and
$R_1$ are not tricritical points. In particular,
this fact is in accordance with the result for pure systems \cite{lyra}.
Also, following \cite{lyra}, we can look on the lines connecting the origin
and fixed points $H_1$ and $R_1$ in Figs. \ref{fig:3} and \ref{fig:4} as 
boundary lines between two second order phase transition regions
with different critical behaviors.

For the constrained case $g_c^{(0)}\not=g_c^{(1)}$ the runaway flow
trajectories intersect the boundary of stability range. The critical
behavior at the boundary of that stability region is determined by the
points $RR_1$ $(n<4)$ and $RH_1$ $(n>4)$.
We suppose that this means rather the smeared phase transition than the phase
transition of first order, in accordance with \cite{lubensky} and 
\cite{aharony}.

In summary, we considered the influence of disordering upon
the critical behavior in systems with hidden degrees of freedom taking into
account the possible role of the constraint upon these degrees of freedom.
It is shown that there are the tricritical fixed points in the constrained
systems and we have rather smeared phase transition than the first order
transition,
while in the unconstrained system all phase transitions are the phase
transition of the second order.

\end{document}